# Design of an Automated Ethanol Vapor Generating System for Alcohol Use Disorder (AUD) Animal Studies

Alexander Pozhitkov[1], Douglas Ramsay[2], Peter A Noble[3]

1: Buddyengineer, Altadena, CA 91001, USA. Email: alexpozhitkov@buddyengineer.com

2: Oral Health Sciences, University of Washington, Seattle, WA 98195 Email: ramsay@uw.com

3: Department of Microbiology, University of Alabama, Birmingham, AL 35294, USA. Email: panoble@uab.edu

## Abstract
Alcohol Use Disorder (AUD) is a prevalent addictive disorder affecting an estimated 29.5 million Americans. It is characterized by impaired control over alcohol consumption despite negative consequences. The number of diagnostic criteria met by an individual typically determines the severity of AUD. Research into AUD focuses on understanding individual susceptibility differences and developing preventive strategies. Alcohol vapor inhalation has emerged as a promising method for pathophysiological investigations in animals, allowing researchers to control the dose and duration of alcohol exposure. This approach is crucial for studying the escalation of voluntary alcohol-drinking behavior. Current commercial systems for alcohol vapor generation have limitations, including combustion risks and the need to adjust multiple parameters. Other methods, like bubbling or blow-over evaporation, face challenges in maintaining equilibrium and avoiding aerosolization. To address these issues, a new type of ethanol vapor generating system is proposed that relies solely on temperature control, creating a vacuum into which ethanol evaporates under thermodynamic control. This approach eliminates the need to adjust multiple parameters and offers improved accuracy and precision in vapor dose delivery. We validated the system as anticipated, achieving stable ethanol vapor after a few priming cycles. Using a 1.2 L cylinder, we obtained approximately 3.6 L of saturated vapor/air mix in 1 minute. Gravimetric results showed that each cycle produced about 100 mg/L or ~5.3% v/v vapor-to-air mixture. The intended use of the ethanol vapor generator is to provide a concentrated ethanol vapor / air mixture to be further diluted before delivering to the animals.

## Introduction

Alcohol Use Disorder (AUD) significantly impacts 11% of Americans, who struggle with an impaired ability to stop or control alcohol use, and the severity of this disease varies from one individual to another. Alcohol vapor inhalation (1,2) is employed in animal models of AUD to study the induction and assessment of alcohol dependence and withdrawal syndrome (3). Vapor inhalation offers distinct advantages over alternative methods, such as forced alcohol administration, due to its non-invasive nature, precise dose control, and the capacity to evaluate withdrawal and abstinence-related behaviors in animals (4). However, the accuracy and precision of vapor dosages are compromised by non-equilibrium conditions and complicated controls (including manual).

The demand for ethanol vapor generating systems in AUD research is significant. A SCOPUS search, conducted on August 22, 2023, using the terms "ethanol vapor animal model", yielded 436 peer-reviewed articles. Ethanol vapors, for example, have been used to investigate cognitive deficits in adult rats resulting from prenatal exposure to inhaled ethanol (5), sex-based differences in behaviors from chronic ethanol exposure (6), the enhancement of voluntary alcohol-drinking behavior linked to the upregulation of dopamine and serotonin (7,8), genetic selection for high alcohol preference and alcohol dependence (9), recovery of orbitofrontal cortex function (10), effects of age and early-life alcohol exposure (11,12,13), genotype matching for alcohol exposure (14), correlation of membrane fluidity with behavioral dependence (15), GABA input plasticity in mice (16), alcohol withdrawal-induced pain in individuals with AUD (17), rest-activity cycle fragmentation during light periods post-withdrawal, and persistent sleep disturbances after alcohol withdrawal (18), alcohol dependence in adult mice and rats following chronic, intermittent ethanol vapor exposure (19), the blockade of orexin receptors preventing alcohol-seeking behavior (20), analysis of metabolomics, mRNA-seq, and microRNA-seq in myocardial tissues after alcohol dependence (21), and alcohol vapor self-administration (22).

A crucial issue is the accuracy and precision of ethanol vapor generating systems employed in AUD research. The precise dose of ethanol vapor received by each individual animal is pivotal to (i) elucidating variations in disease severity that lead to escalated voluntary alcohol-drinking behavior, and (ii) exploring the neurobiological, neurochemical, and neuropharmacological mechanisms that underlie alcohol addiction and withdrawal. Other ethanol vapor generating systems depend on multiple parameters, including many under non-equilibrium conditions. Non-equilibrium conditions tend to be time-dependent, irregular, and changing in unpredictable ways (23), which is problematic for controlling vapor dosage.

To control ethanol vapor dosage, our objective is to design and test a novel automated ethanol vapor-generating system based on thermodynamic principles and equilibrium conditions that yield accurate and precise ethanol vapor concentrations and solely relies on a single parameter: temperature under equilibrium conditions.

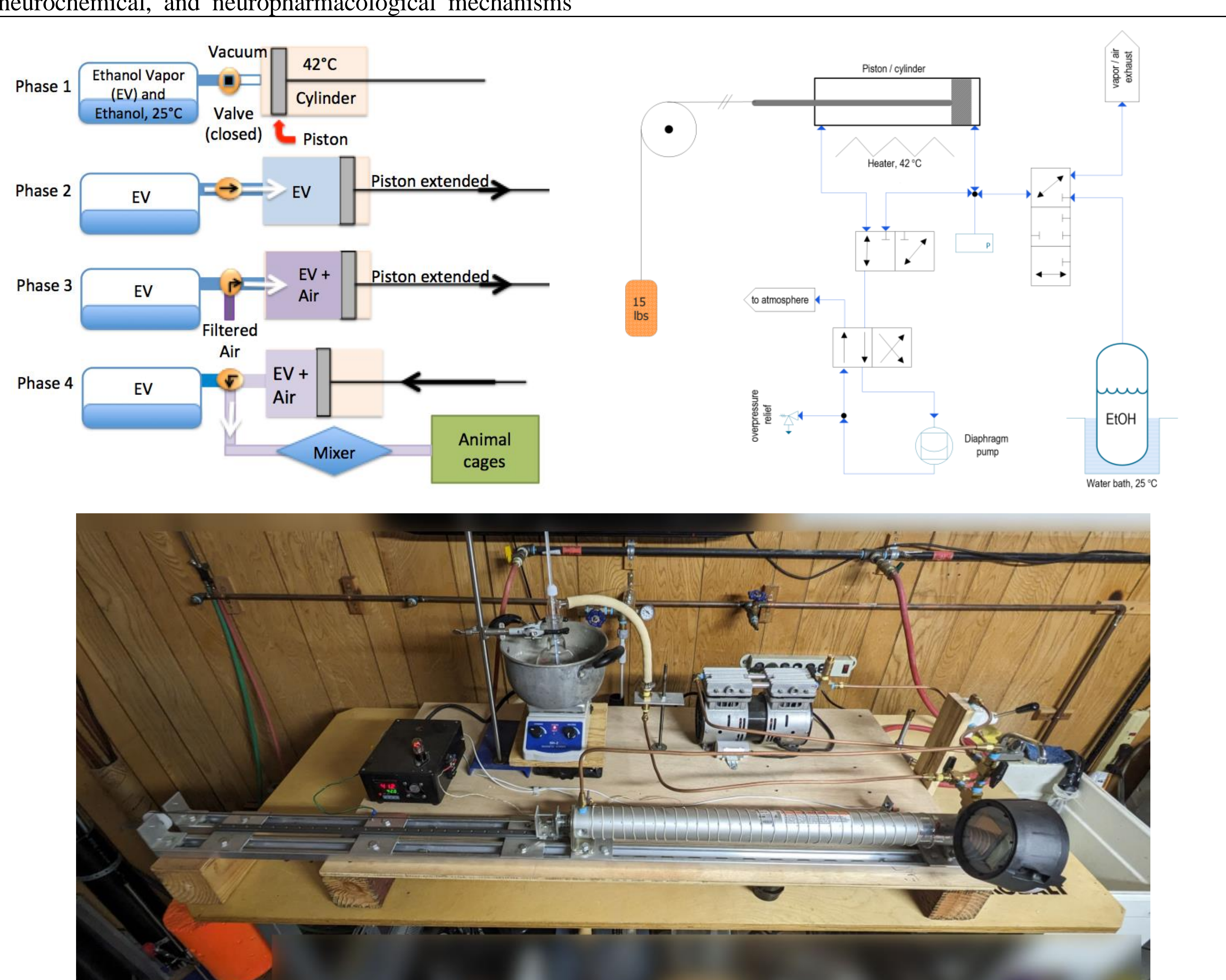

Figure 1. Top left: **An overview showing the relationship between the ethanol flask (25°C), the pneumatic cylinder and piston and the connections to the animal cages. At phase 1, 95% ethanol in the flask is equilibrated with its vapor in the headspace, and the air in the left side of the cylinder is evacuated using the diaphragm pump (not shown). At phase 2, the piston is moved to the right, pressure further drops, and a valve is opened allowing the ethanol vapor in the flask to fill the left side of the cylinder. At phase 3, the air is admitted to the left side of the cylinder. At phase 4, air is pumped into the right side cylinder using the diaphragm pump (not shown) and the piston moves in the left side of the cylinder to export the vapor to the mixer and animal cages. At same time, the valve to the flask is blocked to avert the ingress of air.** Top right: **schematic diagram of ethanol flask (25°C), the pneumatic cylinder and piston and the connections to the animal cages.** Bottom: **actual photograph of the experimental setup.**

# Material and Methods

**Experimental Design.** A simplified overview is depicted in Figure **1**, top left, showing four phases as ethanol vapor flows from the flask through a valve to the left side of the pneumatic cylinder and is then exported to the mixer/animal cages.

*Phase 1*, the 95% ethanol is heated to 25°C and mixed to equilibrate with its vapor in the headspace. The system is initially primed to remove all the air from the evaporation flask (described in section below). Also at phase 1, the air on the left side of the cylinder is evacuated using the diaphragm pump.

*Phase 2*, the pneumatic piston is moved to the right of the cylinder further increasing the vacuum in the left side of the cylinder. At this point, the valve connecting the flask with the ethanol vapor is opened to the cylinder allowing the ethanol vapor to flow into the left side of the cylinder.

*Phase 3*, the air is admitted to the left side of the piston (but not allowed to enter into the evaporation flask).

*Phase 4*, air is then pushed into the right side of the cylinder by the diaphragm pump, forcing the vapor to be exported to the mixer /animal cages. At this point, the ethanol vapor is at ~100% saturation. Before the vapor is expelled, a valve on the flask is blocked to avert the ingress of air. Using 1.2 L cylinder, we obtain ~3.6 L of saturated vapor/air mix in 1 minute.

# Results

**Thermodynamics** Thermal and chemical equilibria occurred at phase 2 (Figure **1**) when the piston is fully extended to the right. For ethanol, the pressure of the saturated vapor in equilibrium with the liquid is determined by the following equation:

$$\log_{10} P = 8.0494 - \frac{1554.3}{222.65 + \mathrm{T}}$$

where $P$ is pressure in mm Hg and $T$ is the temperature in Celsius (25). This equation is relevant to the design of our system because the temperature of the ethanol in the flask is set to 25°C (~7.8 kPa). The temperature is the only parameter controlling the ethanol vapor pressure, hence its quantity in the output. A pneumatic schematic of the design is shown in Figure **1**, top right.

Within the entirety of the system, the pressure aligns with the pressure of saturated ethanol vapors at room temperature by the conclusion of phase 3. Importantly, when the vapor flows from a high-pressure source (the flask) to a low-pressure destination (the cylinder) under adiabatic conditions (no heat exchange with the environment), the Joule Thomson effect leads to a decrease in the temperature of the ethanol vapor. This temperature drop occurs because ethanol vapor does not behave as an ideal gas, following van der Waals' principles. This effect is relevant to our design because the ethanol vapor condenses within the cylinder (experimentally determined).

To prevent condensation, the cylinder is wrapped in a thermal wire and maintained at 42°C. Given the consistent temperature of the ethanol vapor flask at 25°C, coupled with the unchanging temperature of the pneumatic cylinder (held at 42°C), the sole variable influencing the system becomes the temperature of the evaporation flask. During the flow of vapor from the cylinder to the mixer (and eventually to the animal cages), the vapors will attain room temperature without condensation.

**Performance.** Pressure readings were captured using a gauge on the left side of the cylinder, while temperature measurements were taken from the ethanol vapor in the flask using a thermometer. These measurements were obtained at three distinct time points. The first point was recorded in the evacuated cylinder prior to introducing the ethanol vapor (Phase 1). The second point was recorded after the piston is fully extended to the right (Phase 2 before admitting ethanol). Finally, the third point was recorded when the ethanol vapor has been admitted to the cylinder (end of Phase 2).

Two priming cycles during which the air is removed from the flask are shown in blue (Table 1). The following cycles are the production of ethanol vapor. In Cycle 3, Time 1, the pressure was 89.5 kPa below atmospheric pressure following the cylinder's evacuation. At Time 2, the pressure is decreased by 4.5 kPa in relation to Time 1 (= –89.5– (–94.0) ) due to the volume expansion. At Time 3, the pressure had increased by 4 kPa ( = –90.0 – (–94.0) ) in comparison to Time 2 due to admittance of the ethanol vapor.

As the cycles continue, the pressure at Time 3 stays unchanged, so as the pressure differences between Times 2 and 3. This indicates reproducible amount of ethanol evaporated at each cycle solely determined by the flask temperature. It was determined gravimetrically that each cycle produced ~100 mg/L or ~5.3% v/v vapor to air. Our experimental findings suggest that two to three priming cycles are sufficient for air removal from the evaporation flask. The significance of the outcomes shown in Table 1 is the validation of our system operating as anticipated, and stabilization after a few priming cycles.

**Table 1. Temperature and pressure recorded in the left side of the pneumatic cylinder at three time points. Time 1 refers to the pressure in evacuated cylinder before the ethanol vapor is introduced to the cylinder. Time 2 refers to the pressure after the piston is fully extended to the right. Time 3 refers to the pressure when the ethanol vapor has been admitted to the cylinder.**

| Cycle | Temp (°C) | Pressure (kPa) | | |
|---|---|---|---|---|
| | | Time 1 | Time 2 | Time 3 |
| 1 | 23.0 | -89.5 | -94.0 | -78.0 |
| 2 | 23.0 | -89.5 | -94.0 | -89.5 |
| 3 | 23.0 | -88.5 | -94.0 | -90.0 |
| 4 | 23.0 | -89.0 | -93.5 | -90.0 |
| 5 | 23.0 | -88.5 | -93.5 | -90.0 |
| 6 | 23.0 | -88.5 | -93.5 | -90.0 |
| 7 | 23.0 | -88.5 | -93.5 | -90.0 |
| 8 | 23.0 | -88.5 | -93.5 | -90.0 |
| 9 | 23.0 | -88.5 | -93.5 | -90.0 |
| 10 | 23.0 | -88.5 | -93.5 | -90.0 |
| 11 | 23.0 | -88.5 | -93.5 | -90.0 |
| 12 | 23.0 | -88.5 | -93.5 | -90.0 |
| 13 | 23.0 | -88.5 | -93.5 | -90.0 |
| 14 | 23.0 | -88.5 | -93.5 | -90.0 |

# Discussion

Commercial systems for alcohol vapor generation, like the one produced by La Jolla Alcohol Research, involve heating ethanol to create an air-ethanol vapor mix (Figure **2**). Issues include: combustion risk and the need to adjust multiple parameters (airflow rate, dripping rate, and flask temperature), which complicate the accuracy and precision of the dose. Other systems utilize bubbling or blow-over evaporation, both of which have challenges in maintaining equilibrium and avoiding aerosolization. Such systems are difficult to control because bubbling is a highly non-equilibrium process that creates an aerosol of liquid ethanol microdroplets, while blow-over evaporation is affected by the changing height and temperature of the ethanol surface during evaporation. Additionally, variations in temperature between the ethanol surface and the bulk, even after mixing, further complicate the process. Wang et al. (24) required 70 min to attain steady state between ethanol surface and bulk. Having multiple uncontrolled parameters decreases the accuracy and precision of the saturated ethanol vapor delivered to the mixer and ultimately the animal cages.

Our design sets us apart from other systems for alcohol vapor generation because it is based on thermodynamic principles and equilibrium conditions that yield accurate and precise ethanol vapor concentrations and solely relies on a single parameter: temperature under equilibrium conditions.

Considering our success, we present a future designed that can be scale-up alcohol vapor delivery for high throughput laboratories.

1. System Compartmentalization and Re-configuration. The original prototype system will be re-configured from a horizontal to a vertical orientation, which will accommodate two cylinders with pistons adding to the existing single cylinder.
2. Fully Automating the System. We will implement complete automation of the ethanol vapor production system and priming cycle. This involves the use of a programmable logic controller (PLC) to control the solenoid valves. Additionally, the visual pressure gauge will be replaced with a pressure transducer, feeding back to the PLC. The evaporation cycle will be managed through a computer communicating with the PLC. The ethanol reservoir will be isolated from the vapor generator by placing it in a safety cabinet designed to withstand flammable substances. A back up battery will initiate system shutdown upon power outage and send text notifications to end-users.
3. Assess Reproducibility in an AUD Laboratory. We will evaluate the reproducibility of evaporation by measuring final ethanol vapor concentrations (n=10) at reservoir temperatures of 0, 10, 15, 20, 25°C using gravimetric methods and an ethanol vapor sensor. The final vapor concentration is expected to be solely dependent on the reservoir temperature

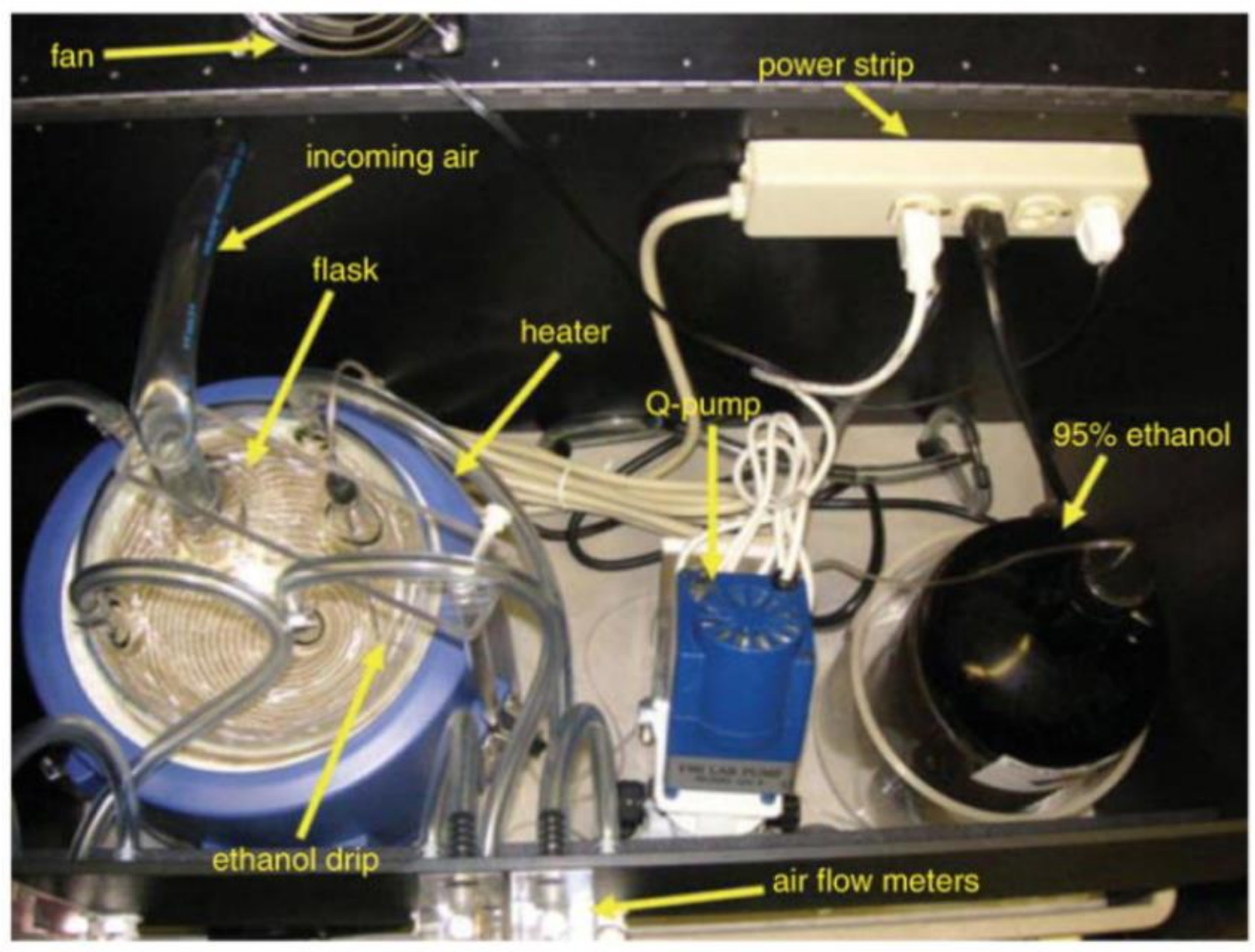

Figure 2. **Aerial view of labeled components housed within the upper shelf of the La Jolla Alcohol Research system. Ethanol is pulled from the 95% ethanol jug by the Q-pump, delivered and dripped into the heated flask, and evaporated there. The vapor is then delivered to housing chambers via plastic tubing and air flow meters. Photo provided from Gilpen et al. (4). The experimenter manually controls the dose, duration, and pattern of exposure of alcohol vapors.**

# Warning

The ethanol–air mixture generated by this system may reach concentrations within the flammable or explosive range (approximately 3–19% v/v) if not properly controlled. Immediately dilute the generated vapor stream to lower, non-flammable concentrations appropriate for animal exposure before it is introduced into exposure chambers or laboratory spaces. Alternatively, by controlling the liquid ethanol temperature, the ethanol–air mixture can be generated directly at a concentration suitable for animal experimentation that remains below the lower flammable limit and does not require further dilution.